\documentclass[10pt,a4paper,twocolumn,english,prl,noshowpacs,noshowkeys,nosuperscriptaddress]{revtex4-1}

\usepackage{mathpazo}
\usepackage[T1]{fontenc}
\usepackage[latin9]{inputenc}
\usepackage{babel}
\usepackage{amsmath}
\usepackage{amssymb}
\usepackage{graphicx}
\usepackage[colorlinks, citecolor=blue, filecolor=black, linkcolor=red, urlcolor=blue]{hyperref}
\usepackage{breakurl}

\begin{document}

\title{Effective equations for repulsive quasi-1D BECs trapped \\
with anharmonic transverse potentials}

\author{Hugo L. C. Couto}

\address{Instituto de F\'{i}sica, Universidade Federal de Goiás, 74.690-900,
Goiânia, Goiás, Brazil}

\author{Ardiley T. Avelar}

\address{Instituto de F\'{i}sica, Universidade Federal de Goiás, 74.690-900,
Goiânia, Goiás, Brazil}

\author{Wesley B. Cardoso}

\address{Instituto de F\'{i}sica, Universidade Federal de Goiás, 74.690-900,
Goiânia, Goiás, Brazil}
\begin{abstract}
One-dimensional nonlinear Schrödinger equations are derived to describe
the axial effective dynamics of cigar-shaped atomic repulsive Bose-Einstein
condensates trapped with anharmonic transverse potentials. The accuracy
of these equations in the perturbative, Thomas-Fermi, and crossover
regimes were verified numerically by comparing the ground-state profiles,
transverse chemical potentials and oscillation patterns with those
results obtained for the full three-dimensional Gross-Pitaevskii equation.
This procedure allows us to derive different patterns of 1D nonlinear
models by the control of the transverse confinement.
\end{abstract}


\maketitle
\emph{Introduction} - It is unanimously recognized the importance
of the experimental realization of Bose-Einstein condensates (BECs)
of atomic dilute gases confined in optical and magnetic traps \cite{Anderson_SCI95,Davis_PRL95},
which have sparked many theoretical and experimental studies of coherent
atomic matter. Since BECs have a long coherence time and can be controlled
and manipulated with enough experimental flexibility by using lasers,
they can constitute highly sensitive sensors for all kinds of force
fields and hold great promise for application to probe magnetic fields
\cite{Vengalattore_PRL07}, to realize high precision inertial navigation
\cite{Zatezalo_IEEE08}, to make Michelson interferometer \cite{Wang_PRL05},
and gyroscope \cite{Gustavson_PRL97}, \emph{etc}.

BECs have also furnished new opportunities to study many-body phenomena
by simulating condensed matter systems in optical lattices \cite{Morsch_RMP06,Bloch_RMP08}
and to investigate nonlinear dynamics of matter waves \cite{Strecker_NAT02}.
Indeed, near zero-temperature BECs can be naturally described by a
mean-field theory in a regime where the system is dilute and weakly
interacting \cite{Pethick_02,Pitaevskii_03}. In this case the system
is ruled by the three-dimensional (3D) Gross-Pitaevskii (GP) equation,
one kind of nonlinear Schrödinger (NLS) equation which admits localized
solutions such as solitons, breathers, and vortices \cite{Malomed_06}.
In particular, the management of the confined profile via optical
lattices and harmonic dipole traps becomes possible to investigate
the effects of dimensionality reduction on the localized solitonic
solution. In fact, the use of a strong trapping in one/two spatial
directions constrains the BEC to assume a disk/cigar-shaped
configuration and obey a quasi 2D/1D dynamics.

It is therefore convenient to develop theoretical models that permit
one to study the condensate dynamics in terms of effective equations
of lower dimensionality taking into account the confinement produced
by highly anisotropic traps. In this regard various approaches have
been developed in recent years \cite{Jackson_PRA98,Chiofalo_PLA00,Salasnich_PRA02,Massignan_PRA03,Kamchatnov_PRA04,Zhang_PRA05,Mateo_PRA08}.
Among them, the effective 1D and 2D nonpolynomial NLS equations by
Salasnich \emph{et al.} \cite{Salasnich_PRA02} and Muñoz-Mateo and
Delgado \cite{Mateo_PRA08} have proven to be the most efficient for
description of BECs with attractive and repulsive interatomic interactions,
respectively. Specifically, in Ref. \cite{Salasnich_PRA02} the authors
used a variational approach to get an effective 1D nonpolynomial NLS
equation by assuming a Gaussian shape for the condensate in the transverse
direction, which is well justified in the limit of weak interatomic
coupling. On the other hand, by applying the standard adiabatic approximation
and using an accurate analytical expression for the corresponding
local chemical potential, the authors of Ref. \cite{Mateo_PRA08}
derived an effective 1D equation that governs the axial dynamics of
mean-field cigar-shaped condensates with repulsive interatomic interactions,
accounting accurately for the contribution from the transverse degrees
of freedom. Following, some theoretical generalizations/applications
for the 1D or 2D reductions were obtained by using the variational
approach in Refs. \cite{Maluckov_PRA08,Gligoric_JPB09,Gligoric_PRA09,Salasnich_PRA09,Salasnich_JPA09,Young_PRA10,Cardoso_PRE11,Salasnich_JPB12,Salasnich_PRA13,Salasnich_PRA14}
and via standard adiabatic approximations in Refs. \cite{Wang_PLA10,Nicolin_PA10,Buitrago_JPB09,Couto_JPB15,Middelkamp_PRA10,Theocharis_PRA10,Mateo_PRA11,Cardoso_PRE13,Mateo_PRE13,Yang_JPB14}.
However, in all these papers the effective equations are obtained
based on the assumption that the transverse potential is quadratic.
In the present letter, we relax this constraint and obtain effective
equations for the longitudinal direction when the transverse potential
is nonquadratric, which opens the possibility of
engineering different types of nonlinearities by control of transverse
potential.

\emph{The model.} - We assume a monoatomic BEC of a dilute atomic
gas, near zero temperature. This system can be accurately described
by the 3D-GPE equation \cite{Pethick_02,Pitaevskii_03}
\begin{equation}
i\hbar\partial_{t}\Psi=\frac{-\hbar^{2}}{2m}\nabla^{2}\Psi+V_{\perp}(\rho)\Psi+V(z)\Psi+g\left|\Psi\right|^{2}\Psi\ ,\label{eq:gpe}
\end{equation}
where $\Psi$ is the normalized density amplitude of the condensation
state, $m$ is the mass of the atomic specie, $\nabla^{2}$ is the
3D Laplacian operator, and $V_{\perp}(\rho)$ and $V(z)$ are respectively
the transverse and the longitudinal parts of a cylindrically symmetric
trap. The nonlinear intensity factor $g=4\pi\hbar^{2}aN/m$ depends
on the s-wave scattering length $a$ and on the number of atoms $N$
in the condensate. When the transverse potential is much more stringent
than the longitudinal one, the characteristic longitudinal time scale
is much lesser than the characteristic transverse one \cite{Jackson_PRA98}.
In this case, the condensate assumes a cigar-shaped form and $\Psi$
can be accurately factorized as a product of the form $\Psi(\mathbf{r},t)\approx\varphi(\rho,n(z,t))\phi(z,t)$,
such that $n(z,t)\equiv\int d^{2}\boldsymbol{\rho}|\Psi|^{2}=|\phi|^{2}$
and $\int dz\,n(z,t)=1$ \cite{Mateo_PRA08}.

The application of this \emph{ansatz} on the 3D-GPE (adiabatic approximation),
followed by the assumption that the characteristic longitudinal length
scale is much greater than the transversal one, results in an 1D evolution
equation to the longitudinal density amplitude $\phi$, given by
\begin{equation}
i\hbar\partial_{t}\phi=-\frac{\hbar^{2}}{2m}\partial_{zz}\phi+V(z)\phi+\mu_{\perp}(n)\phi\ ,\label{eq:psimuperp}
\end{equation}
whose effective interatomic interaction term is defined by $\mu_{\perp}(n)\equiv\int d^{2}\boldsymbol{\rho}\,\varphi^{*}\left(-\frac{\hbar^{2}}{2m}\nabla_{\perp}^{2}\varphi+V_{\perp}\varphi+gn\left|\varphi\right|^{2}\varphi\right)$,
being $\nabla_{\perp}^{2}$ the transverse part of the Laplacian operator.
This quantity, actually the transverse chemical potential, can be
calculated as the lower eigenvalue of the nonlinear eigenvalue problem
\begin{equation}
\mu_{\perp}\varphi=-\frac{\hbar^{2}}{2m}\nabla_{\perp}^{2}\varphi+V_{\perp}\varphi+gn|\varphi|^{2}\varphi\ .\label{eq:muperp}
\end{equation}

Although Eq. \eqref{eq:muperp} cannot be generally solved, it has
well defined limits. In the perturbative regime ($gn\to0$), $\varphi$
is very close to the fundamental state $\varphi_{0}$ of the linear
problem $\mu_{0}\varphi_{0}=-\frac{\hbar^{2}}{2m}\nabla_{\perp}^{2}\varphi_{0}+V_{\perp}\varphi_{0}$.
Taking the substitution $\varphi\to\varphi_{0}$ on Eq. \eqref{eq:muperp},
one obtains the perturbative approximation $\mu_{\text{p}}$ to $\mu_{\perp}$,
given by
\begin{equation}
\mu_{\text{p}}(n)=\mu_{0}+gI_{4}n\ ,\label{eq:nlincub}
\end{equation}
being $\mu_{0}$ the eigenvalue of the linear problem correspondent
to $\varphi_{0}$, and being $I_{4}\equiv\int d^{2}\boldsymbol{\rho}|\varphi_{0}|^{4}$.

In the opposite regime ($gn\to\infty$), one can ignore the kinetic
term and obtains the Thomas-Fermi (TF) approximation
\begin{equation}
\mu_{\text{TF}}\,\varphi_{\text{TF}}=V_{\perp}\varphi_{\mathrm{TF}}+gn|\varphi_{\text{TF}}|^{2}\varphi_{\text{TF}},\label{eq:muperpTF}
\end{equation}
with $\varphi_{\text{\text{TF}}}=\sqrt{\frac{\mu_{\text{\text{TF}}}-V_{\perp}}{gn}}$
everywhere $V_{\perp}(\rho)<\mu_{\text{\text{TF}}}$, and $\varphi_{\text{\text{TF}}}=0$
elsewhere. The normalization of this function results in a relation
between $n$ and $\mu_{\text{TF}}$, namely
\begin{equation}
\pi R_{\text{TF}}^{2}V_{\perp}(R_{\text{TF}})-2\pi\int_{0}^{R_{\text{TF}}}rV_{\perp}(r)dr=gn\ ,\label{eq:nRTF}
\end{equation}
with $R_{\text{TF}}\equiv V_{\perp}^{-1}(\mu_{\text{TF}})$ (assuming
that the transverse potential $V_{\perp}$ has a unique inverse $V_{\perp}^{-1}$).

Between the perturbative and the TF regimes neither $\mu_{\text{p}}$
nor $\mu_{\text{TF}}$ is a good approximation to $\mu_{\perp}$.
As the nonlinearity $\mu_{\perp}(n)$ goes away from the TF regime,
the condensate longitudinal width becomes smaller, and consequently
the trapping contribution to $\mu_{\perp}$. Since the sum of the
kinetic and the trapping contributions to $\mu_{\perp}$cannot be
lesser than $\mu_{0}$, the TF approximation must be conveniently
modified to guarantee this limit. This can be reached by the very
simple substitution of $V_{\perp}$ by $\mu_{0}$ everywhere $V_{\perp}<\mu_{0}$
in Eq. \eqref{eq:muperpTF}, \emph{i. e.},
\begin{equation}
\mu_{\text{i}}\varphi_{\text{i}}=\left\{ \begin{array}{lr}
V_{\perp}\varphi_{\text{i}}+gn|\varphi_{\text{i}}|^{2}\varphi_{\text{i}}\  & (\mu_{0}\le V_{\perp}(\rho)),\\
\mu_{0}\varphi_{\text{i}}+gn|\varphi_{\text{i}}|^{2}\varphi_{\text{i}}\  & (V_{\perp}(\rho)<\mu_{0}).
\end{array}\right.\label{eq:incompleta}
\end{equation}
With this modification we abdicate to know $\varphi$ in the center
of the condensate $(V_{\perp}(\rho)<\mu_{0})$, conjecturing only
a mean value $\varphi_{i}=\sqrt{(\mu_{i}-\mu_{0})/gn}$. The normalization
of $\varphi_{\text{\text{i}}}$ results in
\begin{equation}
\pi R_{\text{i}}^{2}V_{\perp}(R_{\text{i}})-\pi R_{0}^{2}V_{\perp}(R_{0})-2\pi\int_{R_{0}}^{R_{\text{i}}}rV_{\perp}(r)dr=gn\ ,\label{eq:nRMD}
\end{equation}
where we employ the definitions $R_{\text{i}}\equiv V_{\perp}^{-1}(\mu_{\text{i}})$
and $R_{0}\equiv V_{\perp}^{-1}(\mu_{0})$. Note that Eq. \eqref{eq:nRMD}
corresponds to the shift $gn\to gn+\pi R_{0}^{2}V_{\perp}(R_{0})-2\pi\int_{0}^{R_{0}}rV_{\perp}(r)dr$\textcolor{magenta}{{}
}in Eq. \eqref{eq:nRTF}.

The perturbative limit of Eq. \eqref{eq:nRMD} is $\mu_{\text{i}}(n)=\mu_{0}+gn/\pi[V_{\perp}^{-1}(\mu_{0})]^{2}$,
which is not identically equal to Eq. \eqref{eq:nlincub}, as should
be. This feature can be corrected by a second modification on Eq.
\eqref{eq:muperpTF}, this time by using the substitutions $\mu_{\text{i}}\to\kappa\mu$
and $\mu_{0}\to\kappa\mu_{0}$. This operation returns 
\begin{equation}
\kappa\mu\psi=\left\{ \begin{array}{lr}
V_{\perp}\psi+gn|\psi|^{2}\psi & (\kappa\mu_{0}\le V_{\perp}(\rho)),\\
\kappa\mu_{0}\psi+gn|\psi|^{2}\psi & (V_{\perp}(\rho)<\kappa\mu_{0}).
\end{array}\right.\label{eq:muperpnMMD1}
\end{equation}
The normalization of this new approximation results in
\begin{equation}
\pi K^{2}V_{\perp}(K)-\pi K_{0}^{2}V_{\perp}(K_{0})-2\pi\int_{K_{0}}^{K}rV_{\perp}(r)dr=gn\ ,\label{eq:nRnMMD}
\end{equation}
where we used the definitions $K\equiv V_{\perp}^{-1}(\kappa\mu)$
and $K_{0}\equiv V_{\perp}^{-1}(\kappa\mu_{0})$. $\kappa$ is a function
of $n$ that interpolates between the two limiting values: $\kappa_{\text{TF}}=1$
in the TF regime and the solution $\kappa_{\text{p}}$ of 
\begin{equation}
\pi\kappa_{\text{p}}\left[V_{\perp}^{(-1)}(\kappa_{\text{p}}\mu_{0})\right]^{2}I_{4}=1,\label{eq:kappanmmd}
\end{equation}
in the perturbative regime.

Our main goal is to calculate an approximation to $\mu_{\perp}(n)$
reliable in the perturbative and TF regimes, and in the regime between
them (\emph{crossover regime}) by the use of Eq. \eqref{eq:nRnMMD}.
The resulting function $\mu(n)$ will be used as the nonlinearity
for the effective 1D model, Eq. \eqref{eq:psimuperp}, which describes
the longitudinal dynamics of the condensate. However, it is not complete
until we set an appropriate function for $\kappa(n)$.

The transverse chemical potential is equal to the sum of kinetic $\langle T\rangle$,
trapping $\langle V_{\perp}\rangle$, and interaction $\langle N\rangle$
terms, such that, $\eta\equiv\langle T\rangle/(\langle N\rangle+\langle T\rangle)\lesssim1$
in the limit $gn\to0$, and $\eta\gtrsim0$ when $gn\to\infty$. Taking
into account this behavior, $\eta$ can be used to interpolate $\kappa$
between its limiting values, i.e.,
\begin{equation}
\frac{1}{\kappa}=\frac{\eta}{\kappa_{\text{p}}}+\frac{(1-\eta)}{\kappa_{\text{TF}}}=1+\eta\frac{1-\kappa_{\text{p}}}{\kappa_{\text{p}}}\ .\label{eq:interpkappa}
\end{equation}
Since $\eta$ depends on $\varphi$ (through energies), it cannot
be calculated directly, but we can use the interpolation of its limiting
expressions, defined by $\eta_{\text{p}}\equiv\langle T\rangle_{\text{p}}/(\langle N\rangle_{\text{p}}+\langle T\rangle_{\text{p}})$
(perturbative regime) and $\eta_{\text{i}}\equiv\langle T\rangle_{\text{i}}/(\langle N\rangle_{\text{i}}+\langle T\rangle_{\text{i}})$
(TF regime). We know that $\langle T\rangle_{\text{p}}\equiv\frac{-\hbar^{2}}{2m}\int d^{2}\boldsymbol{\rho}\,\varphi_{0}^{*}\nabla_{\perp}^{2}\varphi_{0}$
is the kinetic and $\langle N\rangle_{\text{p}}\equiv gI_{4}n$ is
the interaction energies in perturbative regime. Also, we assume $\langle T\rangle_{\text{i}}=\mu_{\text{i}}-\langle V_{\perp}\rangle_{\text{i}}-\langle N\rangle_{\text{i}}$
as an estimate to the kinetic energy in the TF regime, which is defined
in terms of the trapping $\langle V_{\perp}\rangle_{\text{i}}\equiv\int d^{2}\boldsymbol{\rho}\,V_{\perp}\left|\varphi_{\text{i}}\right|^{2}$
and the interaction energies $\langle N\rangle_{\text{i}}\equiv gn\int d^{2}\boldsymbol{\rho}|\varphi_{\text{i}}|^{4}$.
We stress that terms with subscript $i$ are calculated in terms of
the approximation given by Eq. \eqref{eq:incompleta}. In the same
sense of Eq. \eqref{eq:interpkappa}, by using interpolation
$\eta=\eta\eta_{\text{p}}+(1-\eta)\eta_{\text{i}}$, one obtains
\begin{equation}
\eta=\frac{\eta_{\text{i}}}{1+\eta_{\text{i}}-\eta_{\text{p}}}\ .\label{eq:eta}
\end{equation}
The equations \eqref{eq:interpkappa} and \eqref{eq:eta} determine
the functional dependence of $\kappa$ on $n$ and complete the interpolating
1D model.

\emph{Model for the monomial trapping.} - As an example, let us assume
a transverse confining potential with the form $V_{\perp}(\rho)=U\frac{\rho^{\alpha}}{l^{\alpha}}$,
where $\alpha$, $l$ and $U$ are positive real values. We define
the transverse ($l_{\perp}$) and longitudinal ($l_{z}$) length scales
such that $l_{\perp}^{2}V_{\perp}(l_{\perp})=l_{z}^{2}V(l_{z})=\hbar^{2}/2m$,
and the transverse and longitudinal time units by $\tau_{\perp}=\hbar/V_{\perp}(l_{\perp})$
and $\tau_{z}=\hbar/V(l_{z})$, respectively. The energy units are
merely defined by $V_{\perp}(l_{\perp})$ and $V(l_{z})$. Following,
we use the dimensionless variables $\bar{x}$ and $\hat{x}$, related
to its dimensional counterpart by $x=\bar{x}\,[x]_{z}=\hat{x}\,[x]_{\perp}$,
with $[x]_{z}$ and $[x]_{\perp}$ being the corresponding longitudinal
and transverse dimension units.

Next, by taking into account the transverse monomial trapping, we
calculate the transverse chemical potential approximations (\ref{eq:nlincub}),
(\ref{eq:nRTF}), (\ref{eq:nRMD}), and (\ref{eq:nRnMMD}), obtaining
\begin{subequations}
\begin{gather}
\bar{\mu}_{\text{p}}(\bar{n})=\lambda\hat{\mu}_{0}+\lambda8\pi\bar{a}\hat{I}_{4}N\bar{n}\ ,\label{eq:mp}\\
\bar{\mu}_{\text{TF}}(\bar{n})=\lambda\left[\left(\frac{\alpha+2}{\alpha}\right)8\bar{a}N\bar{n}\right]^{\frac{\alpha}{\alpha+2}}\ ,\label{eq:mtf}\\
\bar{\mu}_{\text{i}}(\bar{n})=\lambda\left[\hat{\mu}_{0}^{\frac{\alpha+2}{\alpha}}+\left(\frac{\alpha+2}{\alpha}\right)8\bar{a}N\bar{n}\right]^{\frac{\alpha}{\alpha+2}}\ ,\label{eq:mi}\\
\bar{\mu}(\bar{n})=\lambda\left[\hat{\mu}_{0}^{\frac{\alpha+2}{\alpha}}+\left(\frac{\alpha+2}{\alpha}\right)\frac{8\bar{a}N\bar{n}}{\kappa^{\frac{\alpha+2}{\alpha}}}\right]^{\frac{\alpha}{\alpha+2}}\ ,\label{eq:mour}
\end{gather}
\end{subequations} with $\ensuremath{\lambda\equiv V_{\perp}(l_{\perp})/V(l_{z})=l_{z}^{2}/l_{\perp}^{2}}$.
Note that the expressions \eqref{eq:mp}-\eqref{eq:mour} depict the
nonlinearity in Eq. \eqref{eq:psimuperp}, with $\bar{n}$ being the
local density $|\bar{\phi}(\bar{z},\bar{t})|^{2}$. In this example,
our main result is the Eq. \eqref{eq:mour}, which interpolates the
transverse chemical potential between its limiting expressions \ref{eq:mp}
and \ref{eq:mtf} and still promises to be a good approximation in
the crossover region. We stress that in the harmonic potential case,
$\alpha=2$, the Eq. \eqref{eq:mour} becomes that obtained in Refs.
\cite{Gerbier_EPL04,Mateo_PRA07}. In this sense, our proposal generalizes
the results of Refs. \cite{Gerbier_EPL04,Mateo_PRA07,Mateo_PRA08,Munoz-Mateo_AP09}. 

We still need to obtain the value of $\kappa$. For the present trapping,
the kinetic and interaction energy contributions are, respectively,
given by
\begin{gather*}
\langle\hat{T}\rangle_{\text{i}}=\left(\frac{\alpha}{\alpha+2}\right)\frac{\hat{\mu}_{0}^{\frac{2\alpha+2}{\alpha}}}{8\bar{a}N\bar{n}}\left(L(\bar{n})^{\frac{\alpha}{\alpha+2}}-1\right)\ ,\\
\langle\hat{N}\rangle_{\text{i}}=\frac{\alpha\hat{\mu}_{0}^{\frac{2\alpha+2}{\alpha}}}{8\bar{a}N\bar{n}}\frac{\alpha L(\bar{n})^{\frac{2\alpha+2}{\alpha+2}}-2(\alpha+1)L(\bar{n})^{\frac{\alpha}{\alpha+2}}+(\alpha+2)}{(\alpha+1)(\alpha+2)}\ ,
\end{gather*}
where $L(\bar{n})\equiv1+8\bar{a}N\bar{n}\left(\frac{\alpha+2}{\alpha}\right)/\hat{\mu}_{0}^{\frac{\alpha+2}{\alpha}}$
and $\kappa_{\text{p}}^{-1}=\left[\pi(\hat{\mu}_{0})^{2/\alpha}\hat{I}_{4}\right]^{\frac{\alpha}{\alpha+2}}$.

By the way, as a limit case, one can assume the BEC transversely confined
in a cylindrical box potential by applying $\alpha\to\infty$ to the
monomial potential, such that $V_{\perp}(\rho)=0$ when $\rho<l$,
and $V_{\perp}(\rho)\to\infty$ otherwise. By replacing this limit
to the above expressions, one obtains $\bar{\mu}_{\text{p}}$ unchanged
(see Eq. \eqref{eq:mp}) and
\begin{gather*}
\bar{\mu}_{\text{TF}}(\bar{n})=8\lambda\bar{a}N\bar{n}\ ,\\
\bar{\mu}_{\text{i}}(\bar{n})=\lambda\left(\hat{\mu}_{0}+8\bar{a}N\bar{n}\right)\ ,\\
\bar{\mu}(\bar{n})=\kappa^{-1}\lambda\left(\kappa\hat{\mu}_{0}+8\bar{a}N\bar{n}\right)\ ,
\end{gather*}
with $\kappa_{\text{p}}^{-1}=\pi\hat{I}_{4}$ and $\eta_{\text{i}}=\lambda\hat{\mu}_{0}/\bar{\mu}_{\text{i}}$.
In addition, $\hat{\varphi}_{0}(0\leq\hat{\rho}\leq1)=J_{0}(j_{0,1}\hat{\rho})/\sqrt{\pi}\,J_{1}(j_{0,1})$
and zero otherwise, $\hat{\mu}_{0}=\langle\hat{T}\rangle_{\text{p}}=j_{0,1}^{2}$,
where $J_{0}$ and $J_{1}$ are the zero and first order Bessel functions,
respectively, and $j_{0,1}$ is the first zero of $J_{0}$.

\emph{Results of numerical simulations.} - In view to calculate $\bar{\mu}$,
one needs at first calculate the ground state of the linear 2D problem
$i\partial_{\hat{t}}\hat{\varphi}=-\nabla_{\hat{\perp}}^{2}\hat{\varphi}+\hat{V}_{\perp}(\hat{\rho})\hat{\varphi}$,
its eigenvalue $\hat{\mu}_{0}$, and $\hat{I}_{4}$. Then, by using
Eqs. \eqref{eq:kappanmmd} and \eqref{eq:eta} one obtains $\kappa_{\text{p}}$
and $\eta$, whose values are used to determine $\kappa$ (see Eq.
\eqref{eq:interpkappa}). These parameters are substituted into the
solution of Eq. \eqref{eq:nRnMMD} in view to get $\mu$, which determines
the nonlinearity of the effective interpolating 1D-NLSE model
\begin{equation}
i\partial_{\bar{t}}\bar{\phi}=-\partial_{\bar{z}\bar{z}}\bar{\phi}+\bar{V}(\bar{z})\bar{\phi}+\bar{\mu}\left(\bar{n}\right)\bar{\phi}\ .\label{eq:1dnlinReesc}
\end{equation}

To check the accuracy of the interpolating model \eqref{eq:1dnlinReesc},
we compare its ground-state profile and oscillation patterns with
those obtained via the 3D-GPE. All of these calculations were made
for transverse monomial potentials, for different values of $\alpha$,
trapping anisotropy parameter $\lambda=\{10,100,1000\}$, $\bar{a}N$
ranging from $10^{-3}$ to $10^{3}$ and a harmonic longitudinal confinement.
The imaginary time evolutions and direct simulations of the models
were made by using a split-step algorithm with Crank-Nicolson discretization
method.

\begin{figure}[tb]
\centering \includegraphics[width=0.9\columnwidth]{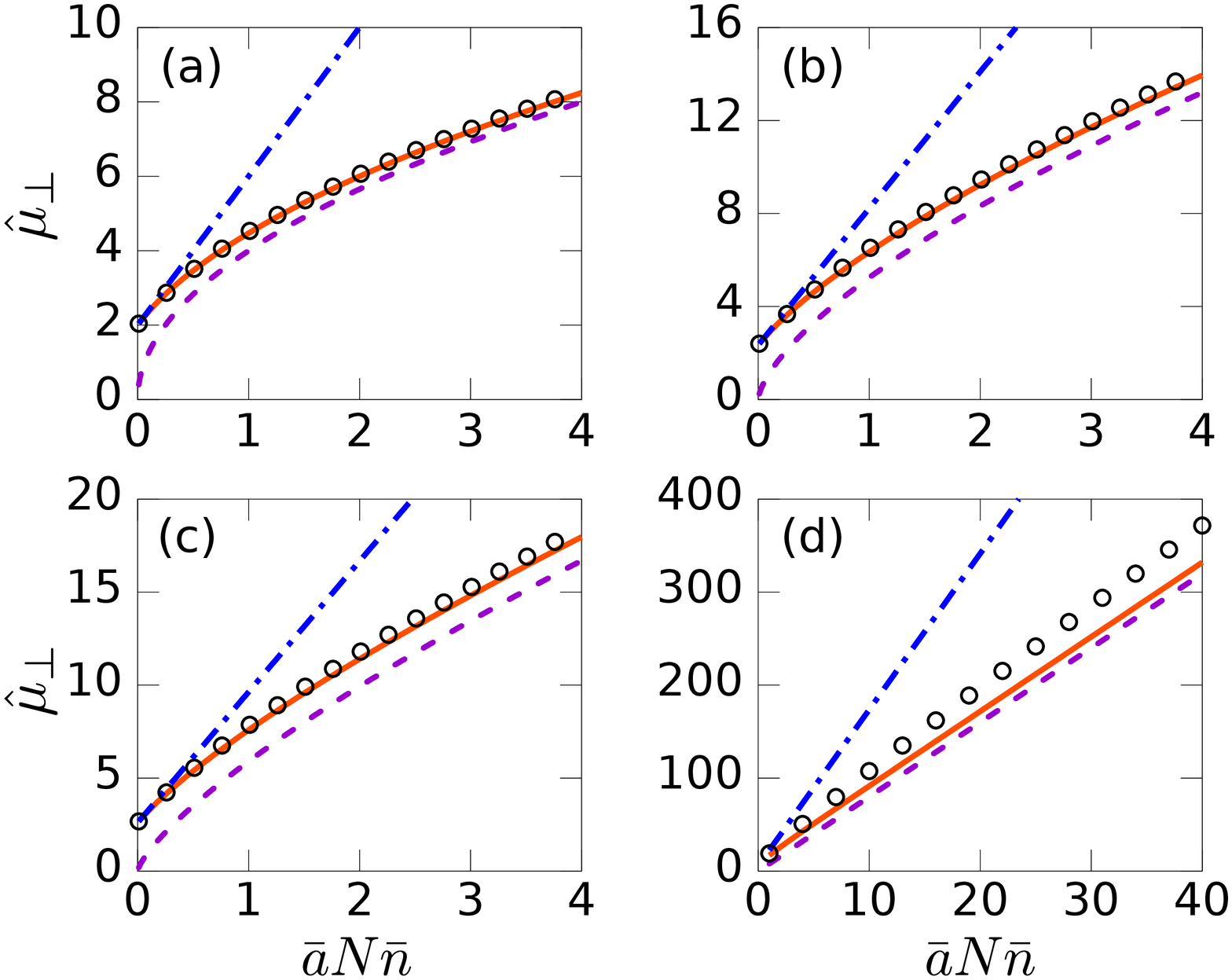} 

\caption{(Color online) Plots of the transverse chemical potentials obtained
by numerical simulations $\hat{\mu}_{\perp}$ in open circles (black),
by the perturbative approximation $\hat{\mu}_{\text{p}}$ in dash-dot
line (blue), by the TF approximation $\hat{\mu}_{\text{TF}}$ in dashed
line (magenta), and by the interpolating approximation $\hat{\mu}$
in solid line (red), as functions of the nonlinearity intensity $\bar{a}N\bar{n}$
for the monomial transverse trapping with (a) $\alpha=2$, (b) $\alpha=4$,
(c) $\alpha=6$, and (d) for the cylindrical box trapping.}

\label{fig:muperp}
\end{figure}

Firstly, by using imaginary time evolution method, we get the ground
state of Eq. \eqref{eq:muperp}, which is used to determine the numerical
value of $\mu_{\perp}$. In Fig. \ref{fig:muperp} we compare this
numerical result with the analytical expressions given by Eqs. \eqref{eq:mp},
\eqref{eq:mtf}, and \eqref{eq:mour}. $\hat{\mu}_{\perp}$ 
is shown as function of $\bar{a}N\bar{n}$ for the monomial trap
potentials with $\alpha=\{2,4,6\}$
and for the cylindrical box trapping ($\alpha\rightarrow\infty$).
In fact, one can note that the transverse chemical potential given
by Eq. \eqref{eq:mour} always presents the best agreement with its
numerical counterpart.

\begin{figure}[tb]
\centering \includegraphics[width=0.9\columnwidth]{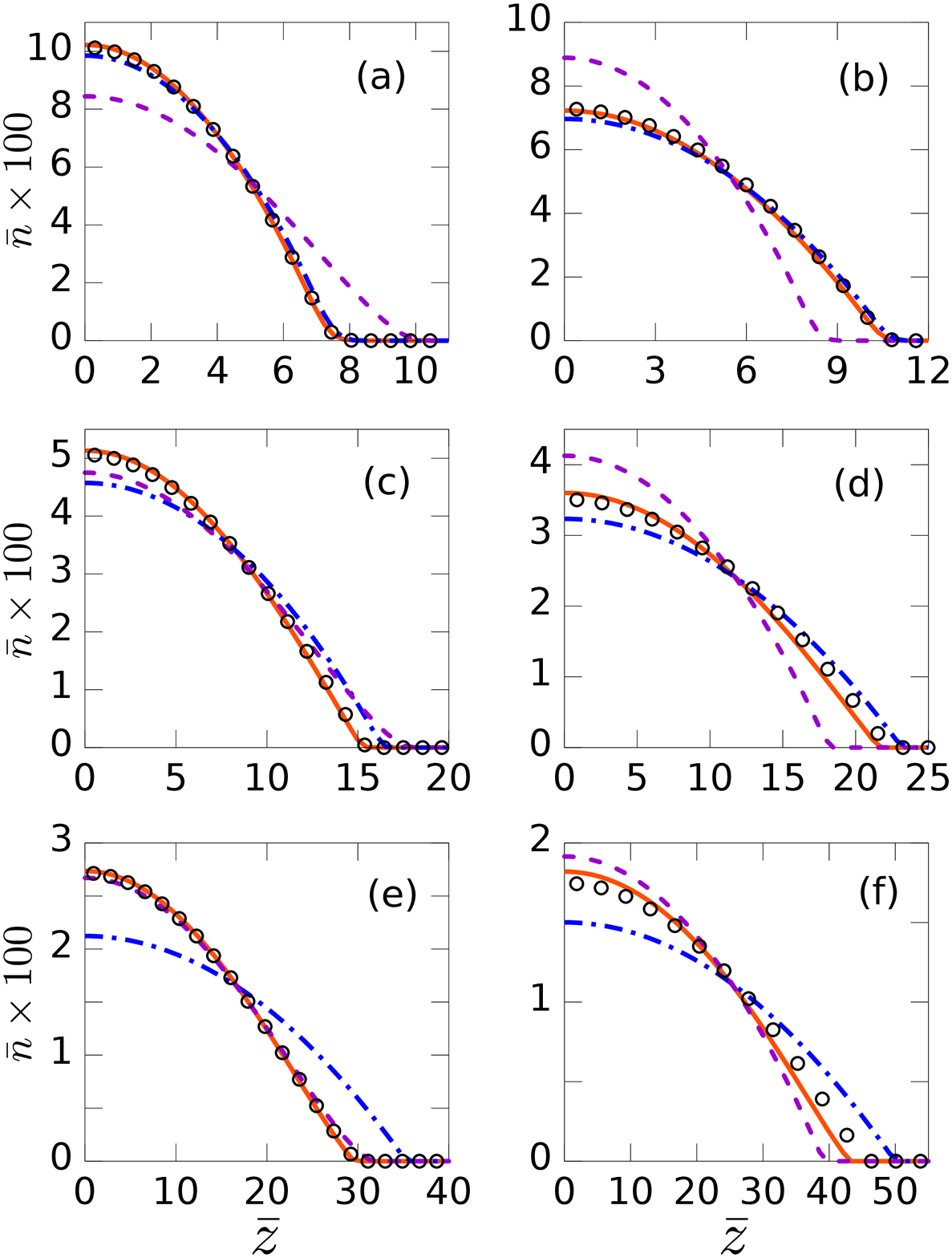} 
\caption{(Color online) Ground state density profiles of 3D-GPE in open circles
(black), 1D perturbative equation in dash-dot line (blue), TF equation
in dash line (magenta), and 1D interpolating equation in solid line
(red) considering a monomial transverse potential ($\alpha=4$) with
(a) $\bar{a}N=1$, (c) $\bar{a}N=10$, and (e) $\bar{a}N=100$, and
considering a cylindrical box transverse potential with (b) $\bar{a}N=1$,
(d) $\bar{a}N=10$, and (f) $\bar{a}N=100$. An axial harmonic confinement
and an anisotropy parameter of $\lambda=100$ were used.}
\label{fig:ground}
\end{figure}

\begin{figure}[tb]
\centering \includegraphics[width=0.9\columnwidth]{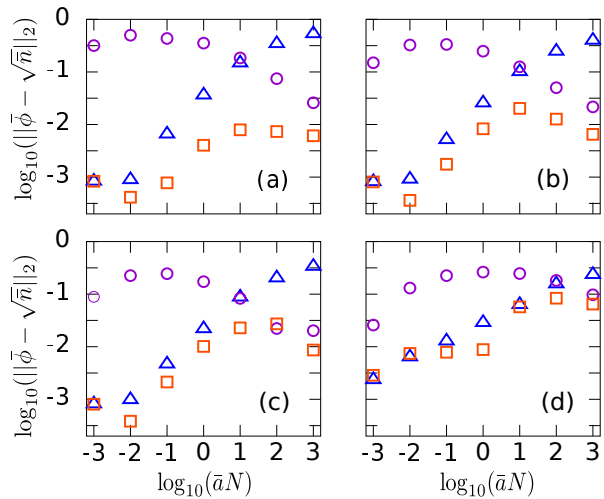}
\caption{(Color online) $L_{2}$-norms ($||\bar{\phi}-\sqrt{\bar{n}}||_{2}$)
with $\bar{\phi}$ being the ground state of the 1D perturbative model
in triangles (blue), the TF model in circles (magenta), and the interpolating
model in boxes (red) and $\bar{n}$ being the 3D-GPE density profile.
We set (a) $\alpha=2$, (b) $\alpha=4$, (c) $\alpha=6$, and (d)
$\alpha\rightarrow\infty$ plus an axial harmonic potential with $\lambda=100$
(anisotropy parameter). }
\label{fig:norm}
\end{figure}

Next, we compare the ground state density distributions of the 1D-NLSEs
in different regimes with its 3D-GPE analog. In Fig. \ref{fig:ground}
we display illustrative examples of the ground state densities obtained
for the monomial transverse potentials with $\alpha=4$ and $\alpha\rightarrow\infty$,
and anisotropy parameter $\lambda=100$. We observe that for $\bar{a}N=1$
the perturbative and the interpolating profiles are close to that
obtained via 3D-GPE, while the TF profile is not so good. Also, by
increasing the value of $\bar{a}N$ to $100$, the TF and interpolating
profiles approach to the 3D-GPE profile while the perturbative profile
departs from it. However, in the intermediate region, $\bar{a}N=10$,
only the interpolating profile fits that from the
3D-GPE. We emphasize that this behavior is general, i.e., we observed
it for all values tested in the range $\bar{a}N\in[10^{-3},10^{3}]$,
$\lambda\in[10,10^{3}]$, and $\alpha\geq2$. Indeed, in view to make
a numerical comparison of the profiles, we display in Fig. \ref{fig:norm}
the usual $L_{2}$-norm $||\bar{\phi}-\sqrt{\bar{n}}||_{2}=\sqrt{\int(\bar{\phi}-\sqrt{\bar{n}})^{2}d\bar{z}}$
in $\log$ scale, where $\bar{\phi}$ is the ground state for each
approximation method and $\bar{n}$ is the 3D-GPE profile. In this
figure we used $\lambda=100$ and four different monomial transverse
traps ($\alpha=\{2,4,6,\infty\}$). By this figure, in agreement with
the visual interpretation of Fig. \ref{fig:ground}, one can note
that the interpolating model presents the lowest values for the $L_{2}$-norm.

\begin{figure}[tb]
\centering \includegraphics[width=0.9\columnwidth]{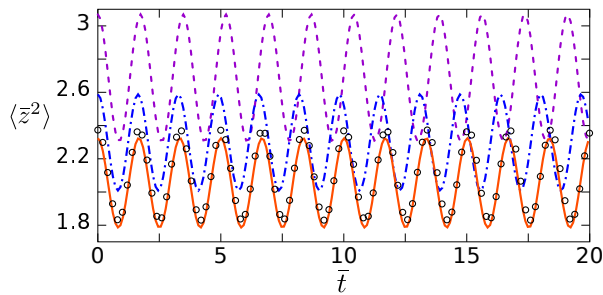} 
\caption{(Color online) Dynamical evolution of axial mean width $\langle\bar{z}^{2}\rangle$
by the 3D-GPE in open circles (black), the perturbative model in dash-dot
line (blue), TF model in dashed line (magenta), and interpolating
model in solid line (red). Here, we consider an axial harmonic potential,
a monomial transversal trap with $\alpha=4$, $\lambda=10$, and $\bar{a}N=1$. }

\label{fig:evol}
\end{figure}

\begin{figure}[tb]
\centering \includegraphics[width=0.9\columnwidth]{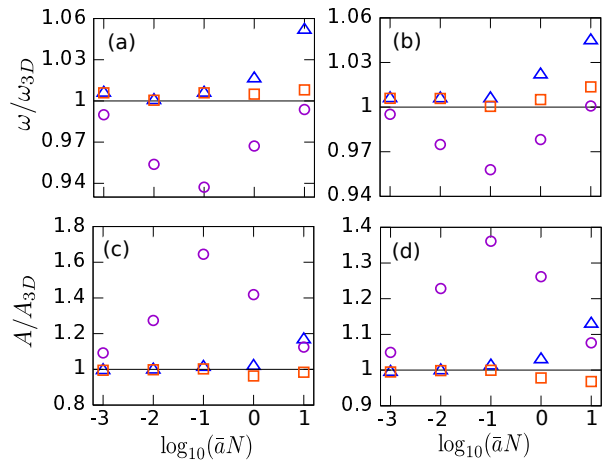} 
\caption{(Color online) Frequencies of oscillation $\omega$ and the corresponding
amplitudes $A$ of $\langle\bar{z}^{2}\rangle$ obtained by the 1D
perturbative model in triangles (blue), the TF model in circles (magenta),
and the interpolating model in boxes (red) with relation to $\omega_{3D}$
and $A_{3D}$, respectively, obtained via 3D-GPE. We set a monomial
transverse trapping $\alpha=4$ (a,c), and a cylindrical box transverse
trapping (b,d). The axial harmonic trapping was changed by $\Omega\to1.1\Omega$.}

\label{fig:wA}
\end{figure}

Finally, we study the dynamical accuracy of the interpolating model
by changing the longitudinal trapping. To this end, we consider the
evolution of the ground state of a BEC, previously confined by a longitudinal
harmonic potential $V(z)=m\Omega^{2}z^{2}/2$, in a new little tighter
harmonic trap, obtained by the replacement $\Omega\to1.1\Omega$.
Indeed, this new trap promotes a pulsation of the condensate axial
profile, as one can see in the axial mean width $\langle\bar{z}^{2}\rangle\equiv\int\bar{z}^{2}|\bar{\phi}|^{2}d\bar{z}$
displayed in Fig. \ref{fig:evol}. Also, by the Fourier analysis of
$\langle\bar{z}^{2}\rangle(t)$, we found the principal frequencies
and amplitudes of oscillation. Comparisons of these quantities by
considering $\alpha=4$ and $\alpha\to\infty$ both with $\lambda=10$
are shown in Fig. \ref{fig:wA}. Note that, differently from the perturbative
and TF models, the frequencies of oscillation $\omega$ and the corresponding
amplitudes $A$ of the interpolating model match the 3D-GPE frequencies
$\omega_{3D}$ and amplitudes $A_{3D}$, respectively, for all nonlinearity
values. We stress that we tested the results shown in Fig. \ref{fig:wA}
for several values of $\alpha$ and $\lambda$, corroborating the
best agreement obtained by the interpolating model as a general feature.

\emph{Conclusion.} - In this letter we derive effective 1D NLS equations
describing the axial effective dynamics of cigar-shaped atomic repulsive
Bose-Einstein condensates trapped with anharmonic transverse potentials.
In this sense, we implemented a modification on the TF approximation
of the transverse chemical potential, which enable us to get accurate
estimates to the ground-state profiles, transverse chemical potentials
and oscillation patterns. Indeed, by numerical simulations we found
that the proposed interpolating model predicts
reasonable values even in the crossover regime (between the perturbative
and TF regimes), being (to the best of our knowledge) the best 1D
model that describes BECs confined by anharmonic transverse potentials.
Although we concentrate on the transverse monomial potential, the
method may be applied to any monotonic transverse potential for which
the TF approximation to $\mu_{\perp}$ is prone to be obtained, offering
no additional difficulties. 
\begin{acknowledgments}
We thank the CNPq (Grant \# 458889/2014-8), FAPEG, and \emph{Instituto
Nacional de Ciência e Tecnologia de Informação Quântica} (INCT-IQ),
Brazilian agencies, for the partial support.

\end{acknowledgments}

\bibliographystyle{apsrev4-1}
\bibliography{Refs}

\end{document}